\documentclass[epj-spec]{svjour}
\usepackage{graphics}
\begin{document}
\title{Benchmark ultra-cool dwarfs in widely separated binary systems}
\author{Z.H. Zhang\inst{1}\fnmsep\thanks{\email{zenghuazhang@hotmail.com}}, D. J. Pinfield\inst{1},  A. C. Day-Jones\inst{2}, B. Burningham\inst{1}, H. R. A. Jones\inst{1}}
\institute{Centre for Astrophysics Research, University of Hertfordshrie, Hatfield AL10 9AB, UK \and Department of Astronomy, Universidad de Chile, Casilla postal 36D, Santiago, Chile}
\abstract{
Ultra-cool dwarfs as wide companions to subgiants, giants, white dwarfs 
and main sequence stars can be very good \emph{benchmark} objects, for which 
we can infer physical properties with minimal reference to theoretical 
models, through association with the primary stars. We have searched for 
benchmark ultra-cool dwarfs in widely separated binary systems using SDSS, 
UKIDSS, and 2MASS. We then estimate 
spectral types using SDSS spectroscopy and multi-band colors, place 
constraints on distance, and perform proper motions calculations for all 
candidates which have sufficient epoch baseline coverage. Analysis of the 
proper motion and distance constraints show that eight of our ultra-cool 
dwarfs are members of widely separated binary systems. Another L3.5 dwarf, SDSS 0832, is shown to be a companion to the bright K3 giant 
$\eta$ Cancri. Such primaries can provide age and metallicity constraints for 
any companion objects, yielding excellent benchmark objects. This is 
the first wide ultra-cool dwarf + giant binary system identified.  
} 
\maketitle
\section{Introduction}
\label{intro}
Current brown dwarf (BD) models have difficulty in accurately reproducing
observations, and benchmark objects are thus needed to calibrate both
atmospheric and evolutionary models. In order for a brown dwarf to be
considered a benchmark it must have one or more properties (e.g. age,
mass, distance, metallicity) that can be constrained relatively independently of
models. BDs as members of binaries are very useful for this purpose, where
the host star can provide age, distance and in some cases
metallicity constraints. How useful these systems are, is dependent on how
well we understand the nature and physics of the primary stars.
Evolved stars, e.g. subgiants, giants and white dwarfs can constraint accurate ages, as well
as metallicity in the case of subgiants and early phase giant stars, via
robust spectroscopic models. These types of binary system can populate the full age range of the disk up to 10 Gyrs\cite{pin}. In this paper we report the discovery of nine common proper motion (PM) binary systems with ultra-cool dwarf (UCD) components. One is a benchmark L dwarf companion to the giant star $\eta$ Cancri, and the other eight are late M and  early L dwarf companions to K-M dwarf stars.

\section{Ultra-cool dwarf selection criteria}
\label{sec:1}

\begin{figure}
\resizebox{1\columnwidth}{!}{%
  \includegraphics{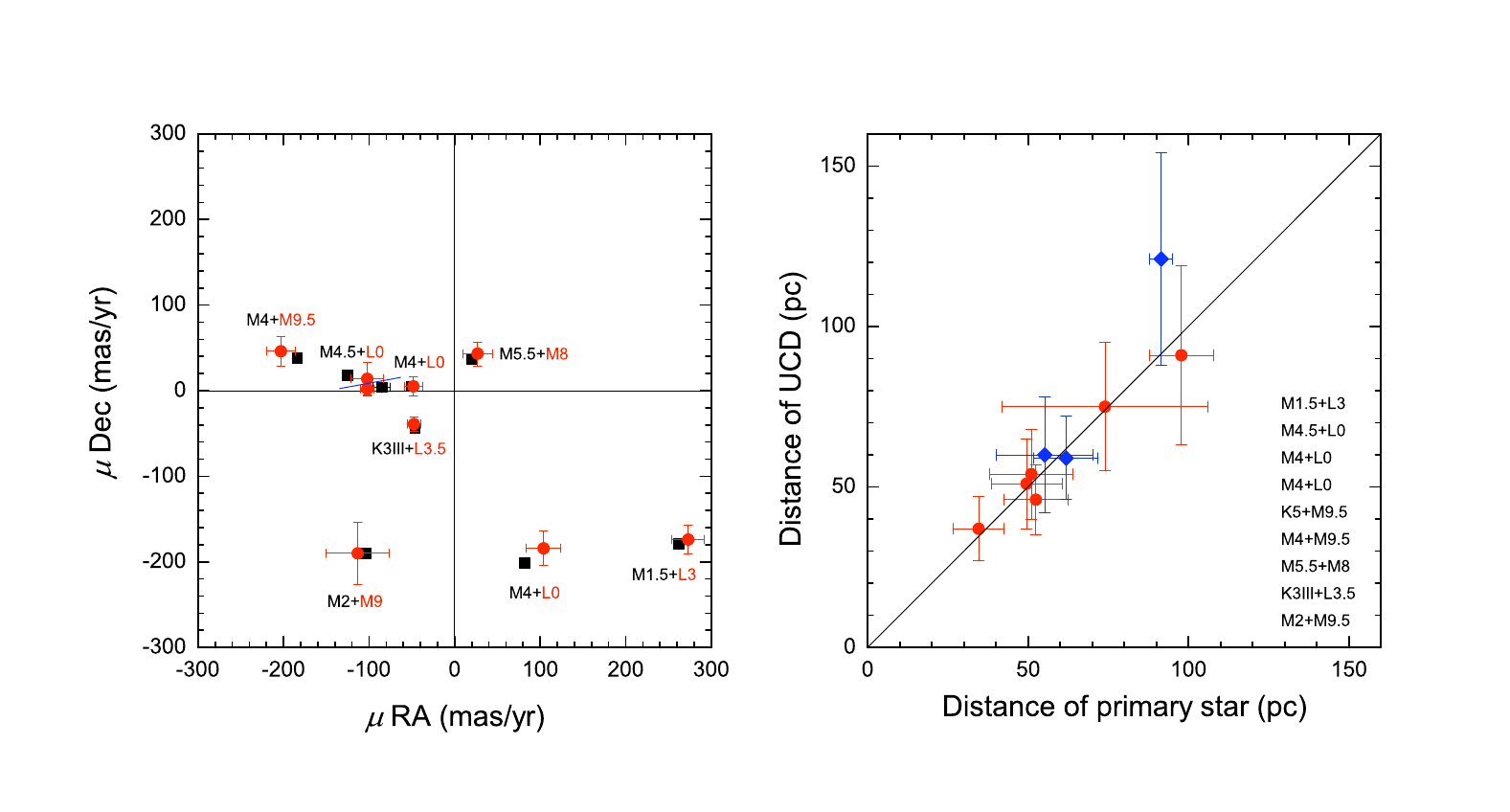} }
\caption{\emph{Left panel}: PMs of nine binary systems. Spectral types of the primary stars (\emph{black squares}) and UCDs (\emph{red circles}) are labeled. A \emph{short line} is used to separated two crowded binary systems.  \emph{Right panel}: Distances of UCDs and their primary star. Spectral type combinations are listed in according distance of primaries in the bottom right of the plot. The \emph{red squares} and \emph{blue diamonds} indicate the objects with and without SDSS spectra respectively.}
\label{pmd9}       
\end{figure}



The  $i-z$ color of the Sloan Digital Sky Survey (SDSS)\cite{aba} as we have shown in our previous work\cite{zha}, is particularly useful for the selection of UCDs. In this work, we focus on a bluer sample to increase the sample size and fraction of objects that have measured SDSS spectra\cite{zhab}.  Our selections here include $1.5<i-z<2$ and $1.5<r-i<4.5$ for objects with SDSS spectra,  and a more constraining selection for objects without spectra, requiring $1.7<i-z<2, 2< r-i<3.3, r-i<7(i-z)-9.3, 16<z<19.5$, and $17<i<21.2$. In addition we placed an additional selection of early L dwarf candidates towards the Praesepe (M44) cluster. We used the same color criteria as previously described for SDSS objects without spectra, but limited our search to $19<z<20.5$, and $21<i<22.4$, corresponding to early L dwarfs at  $\sim$100$-$200 pc.

SDSS selected candidates were cross matched with point sources in the Two Micron All Sky Survey (2MASS)\cite{skr} and  UKIRT Infrared Deep Sky Survey (UKIDSS)\cite{law} catalogues. 563 objects with spectra were cross matched in SDSS and 2MASS, of which 469 were also cross matched in UKIDSS large area survey (LAS). 1761 objects without spectra were cross matched in SDSS and 2MASS, of which 338 objects were cross matched in UKIDSS. So 807 SDSS objects (see online material for these UCDs) were cross matched in 2MASS and UKIDSS of which 469 objects have SDSS spectra. For our UCD sample towards Praesepe, we cross matched candidates with the UKIDSS Galactic Cluster Survey.

For 469 UCD candidates with SDSS spectra, we assigned their spectral type with the {\scriptsize HAMMER} pipeline\cite{cov}. 34 objects of them are new L dwarfs. We used optical-NIR color$-$spectral type relationships\cite{zha,zhb} to assign the spectral types of 338 objects without SDSS spectra. 

\section{Ultra-cool dwarf binaries}
\begin{figure}
\resizebox{0.6\columnwidth}{!}{%
  \includegraphics{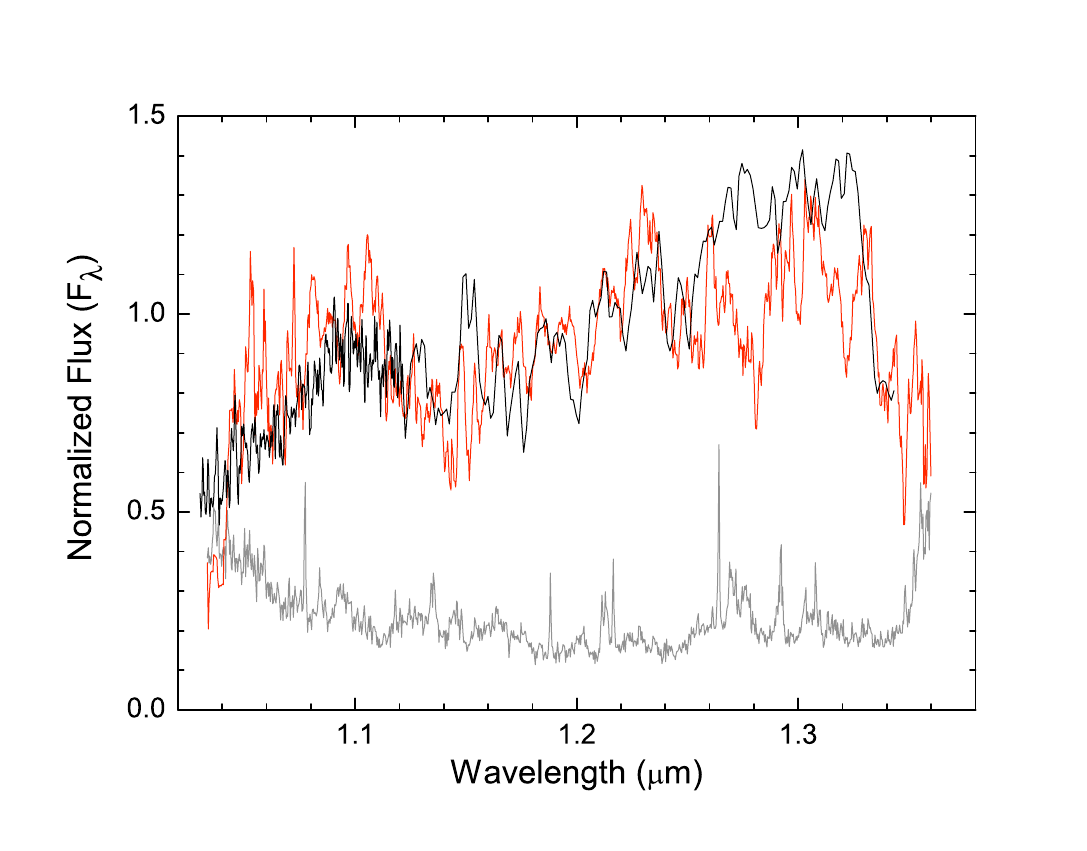} }
\caption{Gemini/NIRI spectrum of $\eta$ Cancri B (\emph{red}). The error spectrum is plotted in \emph{gray}. The spectrum of 2MASS 0028 (\emph{black}) is over ploted, its spectral type in NIR  is L3 and in optical is L4.5.}
\label{nirs}       
\end{figure}

A total of 807 objects in our sample matched in 2MASS-UKIDSS, so we calculated their PMs from their database coordinates and epochs, following the method described in Zhang et al. (2006)\cite{zha}. Eight of them were found to have common PMs to K-M dwarfs. One of our Praesepe candidates, SDSS  J083231.87+202700.0 (SDSS 0832), which was observed with EFOSC2 on NTT, and was found to have a common  PM  with a K3 giant $\eta$ Cancri. The distances of each companion in nine common PM pairs are measured, and distances of companions in each of these pairs are consistent with a binary nature. The statistical probabilities they are random line of sight contamination that have common PMs are all very low ($1.07\times10^{-5}$ by average) have common PMs by random chance are all very low, and we thus conclude that these systems are all genuine associations\cite{zhab}. Figure \ref{pmd9} shows the PMs and distances of these nine binaries. Four of these binaries have projected separations between 300$-$1000 AU, four between 2000$-$3000 AU, and one 15000 AU. Figure \ref{nirs} shows the \emph{J} band spectrum of SDSS 0832. 

%

\begin{figure}
\resizebox{.9 \columnwidth}{!}{%
  \includegraphics{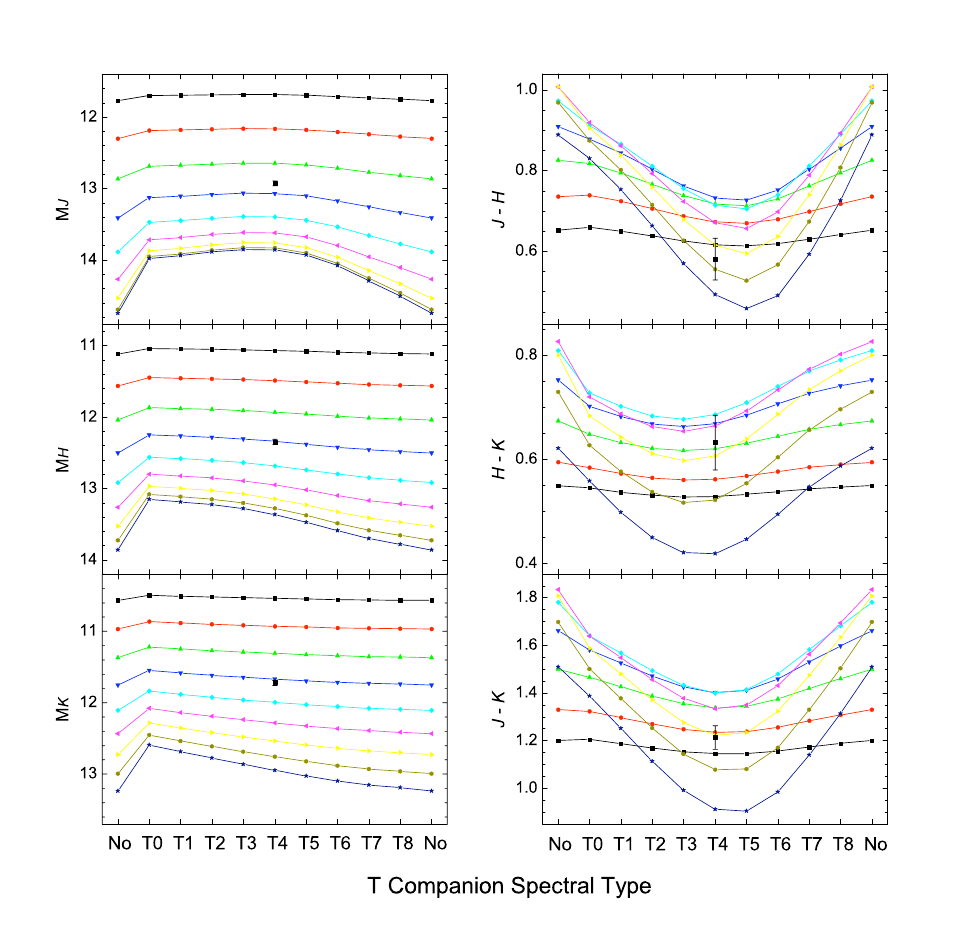} }
\caption{\emph{JHK} absolute magnitudes and colors of unresolved L+T binaries. 
Different \emph{lines} represent absolute magnitudes and colors of L dwarf primaries (L1$-$L9, from top to bottom in left panel) without or with different T companions (T0$-$T8). If $\eta$ Cancri B (\emph{black square}) is actually an unresolved binary, then the constituents would likely be made up of an L4 + a T4 dwarf.}
\label{bmc}       
\end{figure}

\section{Eta Cancri AB}
With a separation of 15019 AU, $\eta$ Cancri AB is an extremely wide binary, however, it is a statistically solid association, with the two components presumably having a common composition and age. 

$\eta$ Cancri A is a bright K3 giant with a \emph{V} band magnitude of 5.33 at a distance of 91.5 pc\cite{van}. It's parameters have been measured by various groups\cite{bro,all,luc,hek}: [Fe/H] = 0.10$\pm$0.08, Mass = 1.50$\pm$0.1 M$_{\odot}$. We have used these properties to constrain the age of $\eta$ Cancri A by comparison to evolutionary tracks. In this way the  evolutionary models\cite{gir} lead to an age constraint of 2.2$-$3.8 Gyrs.

Using the M$_{JHK}-$ spectral type relations\cite{lium} we thus estimate that its spectral type must be L3-4 if it is a single L dwarf. With these magnitudes and an age from $\eta$ Cancri A the evolutionary models\cite{cha00} predict that $M = 68$$-$72 M$_{J}$, $T_{\rm eff}=1920$$\pm$100 K and log ($g/$cms$^{-2})=5.33$$-$5.35. $\eta$ Cancri B has bluer color than normal early L dwarfs. We investigated the possibility that the colors of $\eta$ Cancri B may result from unresolved binarity. T dwarfs have slightly bluer $Y-J$ and significantly bluer $J-H$ and $J-K$ than L dwarfs\cite{kna04,pinf}. Combining the light of an L and T dwarf could thus result in these colors being bluer than those of a single L dwarf. We experimented with a variety of L + T combinations and used the spectral type$-$M$_{JHK}$ relations\cite{lium} to provide absolute magnitudes which we flux-combined to give unresolved color magnitude predictions. We found that an L4 + T4 unresolved combination through SED fitting would have colors and magnitudes very similar to those of $\eta$ Cancri B. Figure \ref{bmc} shows the \emph{JHK} absolute magnitudes and colors of unresolved L+T binaries.

\end{document}